\documentclass[12pt]{iopart}

\usepackage{graphics}
\usepackage{graphicx}
\usepackage{amssymb}
\usepackage{amsthm}
\usepackage{lineno}
\usepackage{rotating}
\usepackage{supertabular}

\newcommand{\sizeimage}{.9}

\begin{document}

\linenumbers

\title[Sample alignment for XRR using CRMs]{Determining sample alignment in X-ray Reflectometry using thickness and density from GaAs/AlAs multilayer certified reference materials}


\author{D Windover$^1$, D L Gil$^1$, Y Azuma$^2$, T Fujimoto$^2$}

\address{$^1$ National Institute of Standards and Technology, Gaithersburg, MD 20899, USA}

\address{$ ^2$ National Metrology Institute of Japan, National Institute of Advanced Industrial Science and Technology, Tsukuba 305-8568, Japan}

\ead{windover@nist.gov}

\begin{abstract}

X-ray reflectometry (XRR) provides researchers and manufacturers with a non-destructive way to determine thickness, roughness, and density of thin films deposited on smooth substrates. Due to the nested nature of equations modeling this phenomenon, the inter-relation between instrument alignment and parameter estimation accuracy is somewhat opaque.  In this study, we intentionally shift incident angle information from a high-quality XRR data set and refine a series of shifted data sets using an identical structural model to assess the effect this angle misalignment has on parameter estimation.  We develop a series of calibration curves relating angle misalignment to variation in layer thickness and density for a multilayer GaAs/AlAs Certified Reference Material on a GaAs substrate. We then test the validity and robustness of several approaches of using known thickness and density parameters from this structure to calibrate instrument alignment.  We find the highest sensitivity to, and linearity with, measurement misalignment from buried AlAs and GaAs layers, in contrast to the surface layers, which show the most variability.  This is a fortuitous result, as buried AlAs and GaAs exhibit the highest long-term stability in thickness.  Therefore, it is indeed possible to use reference thickness estimates to validate XRR angle alignment accuracy.  Buried layer mass density information also shows promise as a robust calibration approach.  This is surprising, as electron density of buried layers is both a highly-correlated phenomenon, and a subtle component within the XRR model.

\end{abstract}


\pacs{61.05.cm, 68.55.jd, 06.20.fb, 07.60.Hv, 06.30.Bp}

\maketitle

\linenumbers

\section{Introduction}

X-ray Reflectivity (XRR) has been used to measure thin films since its discovery by Kiessig in 1931 \cite {kiessig31}.  Parratt in 1954 \cite {parratt54} developed a first-principles approach to modeling XRR data.  Since that time, XRR has matured as the preferred method for non-destructive evaluation of thin film thicknesses (e.g., Lekner, \cite{lekner87} Chason \cite{chason_thin_1997}, and Dalliant \cite{daillant_x-ray_2010}).  XRR patterns result from an interference phenomenon between layers of distinct electron density.  This interference phenomenon manifests as fringes in reflected intensity as a function of the angle of incidence, $\theta$.  The ability of the method to detect interfaces non-destructively is simultaneously its strength and its weakness as a technique:  layers with gradual interfaces, such as layers with inter-diffusion or high roughness, will often be difficult, if not impossible, to model.  Also, because reflection takes place at glancing incidence, the technique illuminates a large sample area during measurement and averages the layer information over this area.  If the sample has lateral thickness or density variation, the interference patterns will also be averaged and can potentially be washed out entirely. However, in the realm of smooth films of high lateral uniformity (e.g., optical coatings and semiconductor structures) XRR has become an invaluable tool.  Both visible light characterization methods (such as optical reflectometry and ellipsometry \cite{kraemer_combined_2010}) and XRR can provide users with thickness and roughness information.  However, the index of refraction of materials in the visible wavelength region often varies dramatically between materials.  This introduces a high degree of uncertainty and modeling challenges with visible light characterization methods.  This problem is eliminated with XRR as the index of refraction, in the hard x-ray spectrum, is near unity for all materials. Further, as the corrections due to index are negligible, XRR thickness information is easily extracted from data, making XRR an ideal approach for International System of Units (SI)-traceable measurements of thickness.  For accuracy in thickness determination, we need calibration artifacts for inter-comparison measurements across XRR tools manufactured and used across the world.


In the last decade, an international collaboration under the Versailles Project on Advanced Materials and Standards (VAMAS) has performed round-robin studies on several candidate materials to be used as thickness standards for XRR \cite{colombi_reproducibility_2008,matyi_international_2008}.  One structure studied by VAMAS, a three-repeat bi-layer of GaAs/AlAs (total of six layers) deposited on a GaAs wafer, was developed by the National Metrology Institute of Japan (NMIJ) as a pre-standard for a NMIJ Certified Reference Material (CRM).  Data taken in 2004 by researchers at NMIJ on a similarly deposited structure, is the focus of this study.  The National Institute of Standards and Technology (NIST) has been developing Bayesian approaches (see Sivia \cite{sivia_data_1996}) to estimate uncertainty in modeling XRR parameters \cite{windover_nist_2007,windover_characterization_2005}.

\section{Experimental Details}

XRR data are collected as a series of incident angle, $\theta^i$, and reflected intensity, $I^R$, data pairs, $(\theta^i_l, I_{l}^R)$ stepped over $N$ points in a range, $l=1, 2, \ldots, N$ from a starting incident angle, $\theta^i_{start}$ to ending incident angle $\theta^i_{end}$.  These data should represent the specular reflection from the surface of the sample.  Specular reflection occurs when the incident and reflected angles are equal:  $\theta^i=\theta^r$.  Our model for XRR starts with the assumption of specular reflection.  However, instrumentation effects, such as sample misalignment, $\delta$, may cause a deviation from this condition.  An XRR measurement instrument does not directly measure $\theta^i$ or $\theta^r$. Instead, a near parallel beam of X-rays impinges on a sample which is rotated to a sample angle, $\omega$.  Intensity data is collected from a detector which is in turn rotated to a a detector angle,  $\theta^d$.  We define an aligned instrument (and sample) to be one where $\omega=0$ implies that $\theta^d=0$.  Specular condition is obtained only when $\omega=\theta^i$ and $\theta^d=2\times\omega=\theta^i+\theta^r$; often called a $\theta - 2\theta$ scan.

The XRR data was collected on a Rigaku\footnote{Certain commercial equipment, instruments, or materials are identified in this poster to specify the experimental procedure adequately. Such identification is not intended to imply recommendation or endorsement by the National Institute of Standards and Technology, nor is it intended to imply that the materials or equipment identified are necessarily the best available for the purpose.} ATX-G type reflectometer configured using a Cu K$_{\alpha}$ rotating anode operating at 50 kV and 300 mA with a graded parabolic multilayer mirror and a Ge (111) channel cut monochromator.  This arrangement produced an intense, highly parallel X-ray source comparable to the brightest commercial XRR instruments available on the market today.  Cu radiation, in laboratory settings, is the most common energy used, and provides the industry baseline for laboratory measurements using X-rays.

XRR sample alignment is difficult to perform, as it requires rocking the sample with the detector at true zero to crudely align the specimen and to set the specimen height. The sample then is roctated at a series of fixed detector angles to find the optimal omega angle for a given (trusted) detector angle. If the detector zero was incorrect, or there are any curvature effects confounding this alignment strategy, or any number of other instrument effects, then the sample and detector angles may be off by a fixed shift, and derived thickness, density, and roughness will be impacted systematically. Angular shifts in XRR intensity information have a pronounced and systematic impacts on the high frequency components of the measured data (film density and thickness). Sample height and edge effects will contribute pronounced off-specular and beam-footprint effects which cause slow varying intensity effects spread across the entire range of the data. These effects can be confounded with surface roughness and the presence or thickness of very thin surface layers.


The GaAs/AlAs multilayer structure used in this study was sample 1-08 of the BAAA4002c series provided by NMIJ (referred to here as 1-08). It was used as a pre-standard in the development of NMIJ Certified Reference Material 5201-a. The GaAs/AlAS layers were deposited using molecular beam epitaxy which produced epitaxial, stoichiometric, and monotonically smooth layers, in this structure. This series of layers were provided by NMIJ to the international community for a long-term study of the stability of this structure. Based on the results of these formal and informal comparisons, NMIJ certified only the thickness of the bottom 4 layers for the final CRM as the upper 2 tended to show instability over time. Density was assumed to be near bulk values, and roughness was assumed to be interatomic-scale (0.3 nm - 0.5 nm).

\section{Theory}

\subsection{Reflectometry modeling}

The first-principles treatment of XRR modeling is discussed in exhaustive detail elsewhere (see \cite{lekner87}).  However, a brief discussion of several key features of the model is useful to understand misalignment impact.  XRR measurement modeling typically uses a Fresnel homogeneous layer model to treat thin films ($j$ = 1...N) in a stack as slabs of a fixed, refractive index, $n_j$.  Refractive index is a complex number with a real component, $\delta_j$, which relates to electron density and an imaginary component, $\beta_j$, which represents absorption of X-rays in a material.  Refractive index is defined by convention as $n_j = 1-\delta_j + i\beta_j$ and the $\delta$ and $\beta$ values for any given element at most energies of interest in characterization are well known.  Parratt developed his analysis method for XRR by solving Snell's Law of refraction for each successive layer in a stack of $N$ layers.\cite{parratt54}  The coefficient of reflection from any layer can be calculated by starting with the perpendicular component of the wave vector from a material layer $j$ and by assuming a fixed $n$ over the layer thickness:

\begin{equation}
k_{zj} = k\sqrt{n_j^2 - \cos^2\theta^i}
\label{eq:kz}
\end{equation}

When an interface is encountered, signifying a new index of refraction $n_{j+1}$, we have reflection of X-rays.  For the interface between the layers $j$ and -$j+1$, we have the following Fresnel relationship, where $r_{j,j+1}$ represents the component of X-rays reflected from the $j+1$ interface back into the $j$ layer:

\begin{equation}
r_{j,j+1} = \frac{k_{zj}-k_{zj+1}}{k_{zj}+k_{zj+1}}
\label{eq:r}
\end{equation}
where wave vector, $k=2\pi/\lambda$, and $\lambda$ is the X-ray wavelength.

Parratt's insight to this problem was to build a recursion relation for the reflection coefficient from every successive interface, $j$ to $j+1$ in a stack, and substitute in the reflections from lower levels of the structure, all the way to the bottom of the stack (i.e., the substrate).

\begin{equation}
X_j = e^{2ik_{zj}z_j}\frac{r_{j,j+1}+ X_{j+1}}{1+r_{j,j+1}X_{j+1}}
\label{eq:X}
\end{equation}

where $z_j$ is the thickness of layer $j$.  




All successive reflection coefficients from $X_{j-1}$  to $X_0$ can be solved by using successive substitutions of Eqns. \ref{eq:r} and \ref{eq:X}, with the special case that, since the substrate is assumed to be infinite, we have no reflection from its lower interface: $X_{N,N+1}=0$ for an N-layer stack. The measured reflection intensity is then $I^R = \left|X_0\right|^2$.

Layer thickness information is ``locked'' into the recursion though each instance of $z_j$, and layer density information enters as a function derived from the index of refraction, $n_j$, for each layer within the recursion. An XRR structural model is defined by a set of parameters for layer thickness $t_j\equiv z_j$, layer density, $\rho_j=f(n_j)$, and layer roughness, $\sigma_j$.  The fundamental observation for this work is that $I^R$ is a function of $\theta^i$ through each instance of $k_{zj}$ in the recursive equation. Separation of these parameters is impossible due to the nested, interdependent nature.  The focus of this study will be to determine both severity and parameter refinement trends given intentional shifting of the observed $\theta^i$ for high quality data and a model test system.

\subsection{Data refinement approach}

Our data refinement approach follows the structure and notation of Wormington \cite{wormington99}.  A model is developed assuming a limited number of layers and a narrow range of allowed (plausible) thickness, roughness, and density parameters which are constrained as tightly as possible by any ancillary measurements and deposition input information.  For example, surface roughness can be measured by atomic force microscopy.  Substrate roughness can be constrained by assuming typical wafer specifications, and cross-sectional thickness information can be constrained using transmission electron microscopy or allowed deposition times and typical deposition rates.  Using this highly constrained structural model, reflection intensity values can be simulated over a range of measurable angles that mimic real measurement data, $(\theta_l, I^{R-calc}_{l})$.  This simulated intensities is then compared to measured intensities, to determine the `goodness-of-fit' for a given set of model parameters, $\textbf{p}$, and this process can be repeated for a range of test cases, $\textbf{p}$.



It is common to show $\chi^2$ or goodness-of-fit (GOF) plots as a function of parameter variation to establish best fit between data and mathematical model, for example see \cite{kolbe_thickness_2005}.  One of Wormington's early achievements in XRR data analysis was the implementation of alternative $\chi^2$ strategies for fitting to data.  XRR measurements often scale over many orders of magnitude requiring modification to traditional data weighting schemes. Wormington implemented several cost functions for XRR data refinement.  The mean square error (MSE) of the log of reflected intensity,  $\log{I_R}$, values provides a robust cost function for refining XRR data and where \textbf{p} represents all the parameters used in the XRR model:

\begin{equation}
MSE_{log}(\textbf{p}) = \frac{1}{N-1}\sum^{N}_{j=1}\left[logI^R_{j} - log I_j^{R-calc} (\theta_j;\textbf{p})\right]^2
\end{equation}

$MSE_{log}$ allows for oscillation data contained several decades below the incident intensity,$I_o$, to still contribute to the model refinement.  Figure \ref{fig:XRR_good_fit} a) shows specular XRR data, with log of reflected intensity, $\log{I^R}$, in counts as a function of incident/reflection angle $\theta$ in radians. Oscillation data is present over nearly seven orders of magnitude in intensity.  Traditional $\chi^2$ would ignore contributions form the last five orders of magnitude of the data.

Wormington's other revolutionary contribution was the use of Differential Evolution (DE), a type of Genetic Algorithm (GA), to refine XRR data.  The GA method tests a large population of potential parameter solutions, $\textbf{p}$, simultaneously against XRR data with one of the modified cost functions, (e.g., $MSE_{log}$).  Of this large population of solutions, only the `fittest'' solutions -- i.e., those with the lowest $MSE_{log}$ -- are allowed to continue.  These solutions are then bred or allowed to intermix parameters $p(x_j)$, and then produce a new large population of solutions for testing against the XRR data.  This culling, breeding, culling model is cycled through hundreds to thousands of generations to select the global minimum solution.  This refinement approach does have one implicit limitation, in that it cannot tell you directly the uncertainty of parameters within a model, but only the best set of parameters.  In this study, we compare these best parameter sets, $\textbf{p}$, for each misalignment and look for variational trends in our parameters.  This tests the sensitivity of each parameter to angular misalignment, provided the GA was successful at finding the global minimum for the misaligned XRR data.  The allowed parameter ranges in a model can sometimes influence the best solution, especially for intentionally misaligned data.  Therefore, we use large ranges for parameter estimates in such a study.

\subsection{Introduction of sample misalignment}

To simulate misalignment of the reflectometer zero angle, we shift the angle, $\theta^i$, by a fixed amount, $\delta$,

\begin{equation*}
\theta^{misalign}_j = (\theta^i_j+\delta),
\end{equation*}

for each angle in the initial, well-aligned dataset. We then generate simulated misaligned data which we then fit for each misalignment, $\delta$ = [-0.025, -0.020, -0.015, -0.010, -0.005, 0, 0.005, 0.010, 0.015, 0.020, 0.025] in $^\circ$. A typical XRR experiment will collect intensity data over an angular range of 3$^\circ$.  A misalignment of 0.005$^\circ$ corresponds to just a 1/600th shift for the total angular range of the scan and as such, may be assumed to have an insignificant effect.  However, our modeling shows that even such a small misalignment can have a substantial impact on parameter estimates.

\section{Results and Discussion}

The initial, well-aligned XRR data-set contained 620 $(\theta_j, I^{R}_{j})$ measurement pairs collected on a commercial XRR instrument at NMIJ using sample 1-08, covering a range of 0$^\circ$ to 3.1$^\circ$ in $\theta_j$  in 0.005$^\circ$ steps.  Each of the 11 data sets, one aligned, and 10 misaligned, was fit using a GA refinement.  Each GA refinement used a population of over 200 members with random parameters, $\textbf{p}$.  Each GA was run for 5,000 generations and from several choices of initial parameter values in order to verify that the true global minimum was found.  Table \ref{table:range} provides the allowed ranges for the 23 model parameters used in our refinements.  Note that our prior knowledge of film thickness information was used to constrain multilayer thickness within a narrow window (1 nm for buried layers), and roughness to atomic feature scales (0.2 nm to 0.5 nm) assuming very-high quality interfaces with no interdiffusion; however, density was allowed to vary across a large range (0.5 to 2 times bulk density values). A second study was performed with a wider range of allowed roughness parameters (Table \ref{table:range_wide}) to address issues in refinement of surface layers for highly misaligned XRR data.  We performed GA refinement with both NIST-developed software\cite{windover_nist_2007,windover_characterization_2005} and a commercial package for comparison (Bede REFS 4.5).  Results from both software packages and both parameter ranges are shown.

\begin{figure}
  \begin{tabular} {cc}
    a)&\resizebox{\sizeimage\textwidth}{!}
  	{\includegraphics{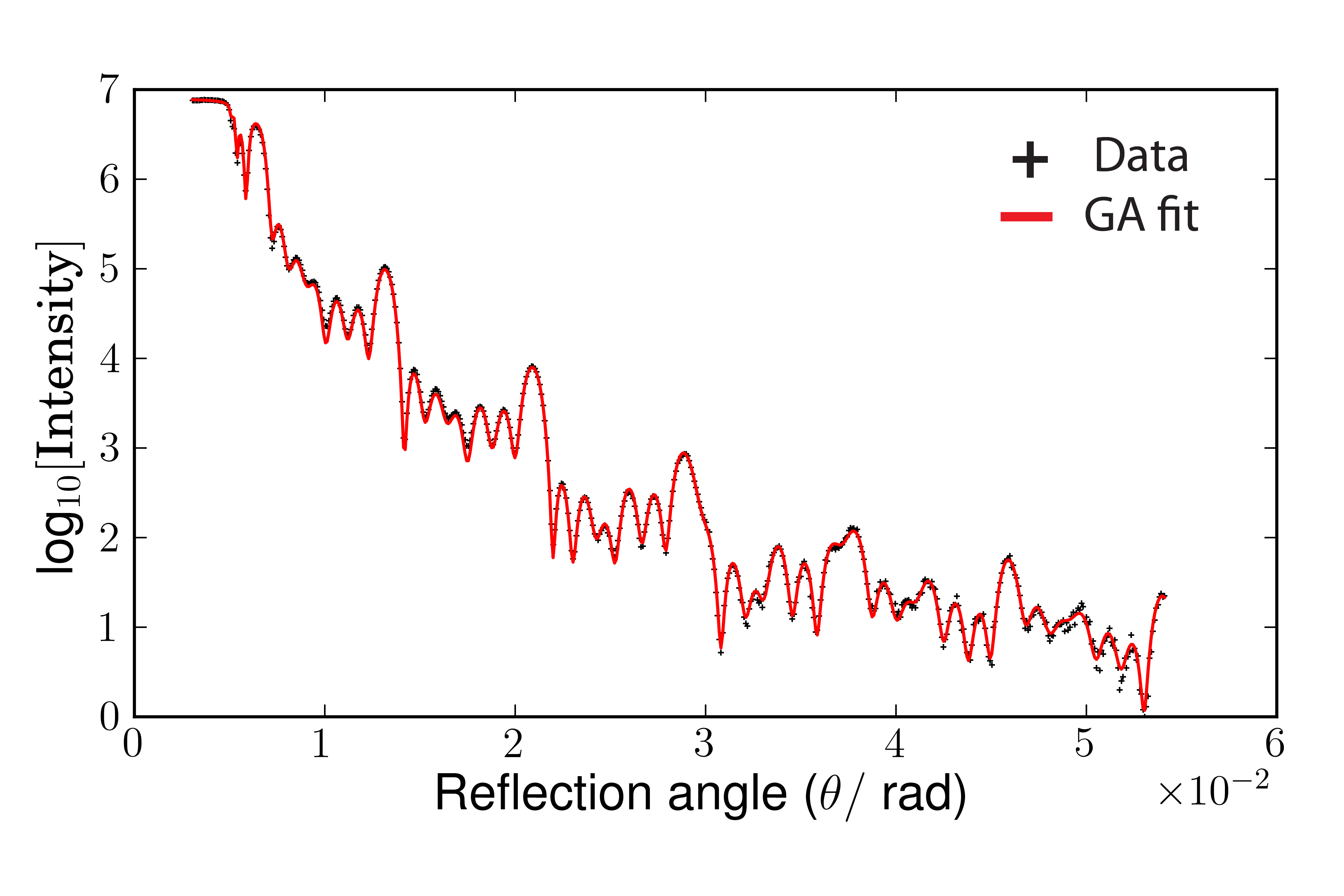}}\\
    b)&\resizebox{\sizeimage\textwidth}{!}
  	{\includegraphics{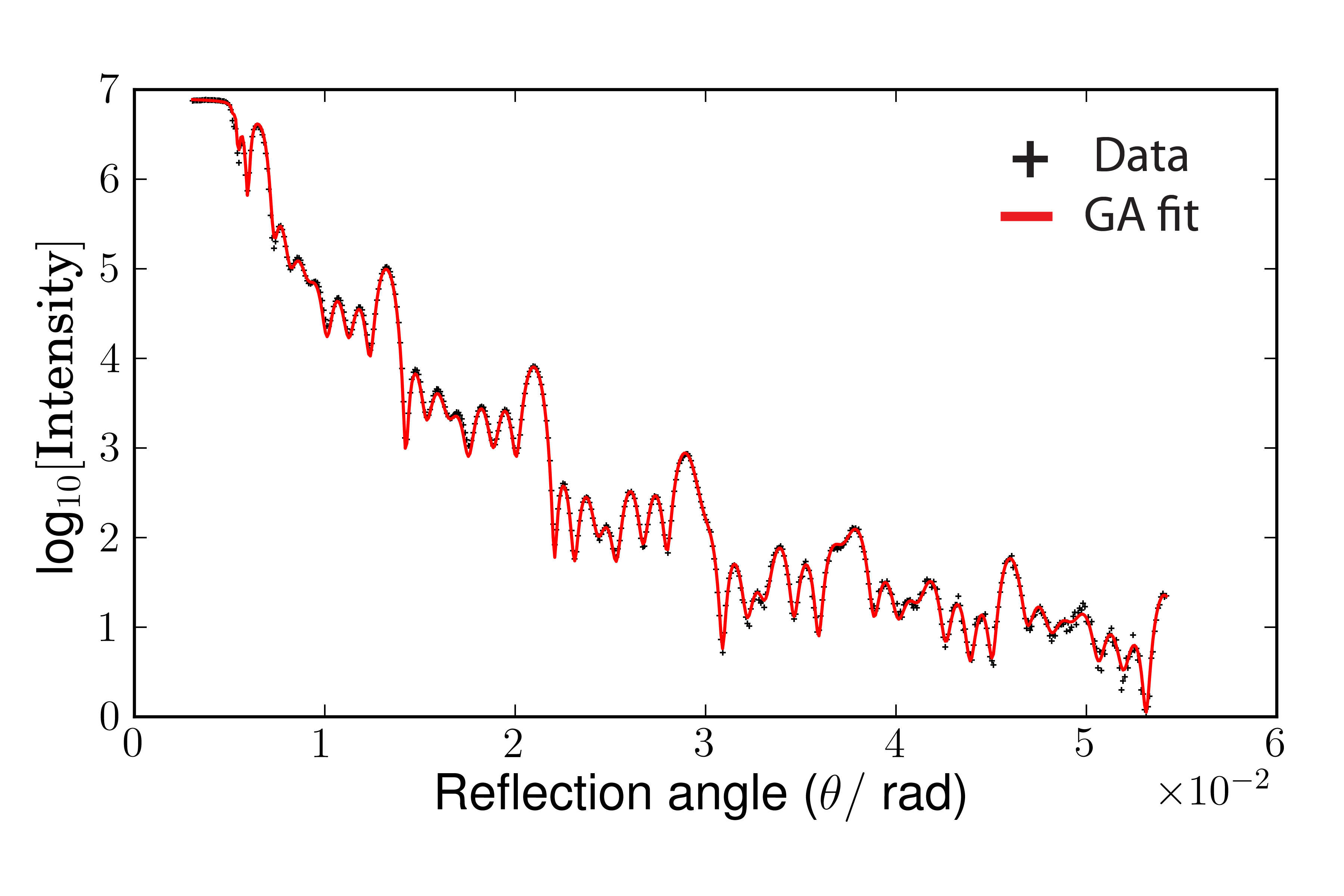}}
  \end{tabular}
  \caption{XRR measurement data and GA model refinement result for:  a) aligned data and b) data intentionally shifted +0.005$^\circ$ $\delta$ in $\omega$.  Data is represented by points with a line representing GA refinement.}
  \label{fig:XRR_good_fit}
\end{figure}

\begin{figure}
  \begin{tabular}{cc}
  a)&\resizebox{\sizeimage\textwidth}{!}
  	{\includegraphics{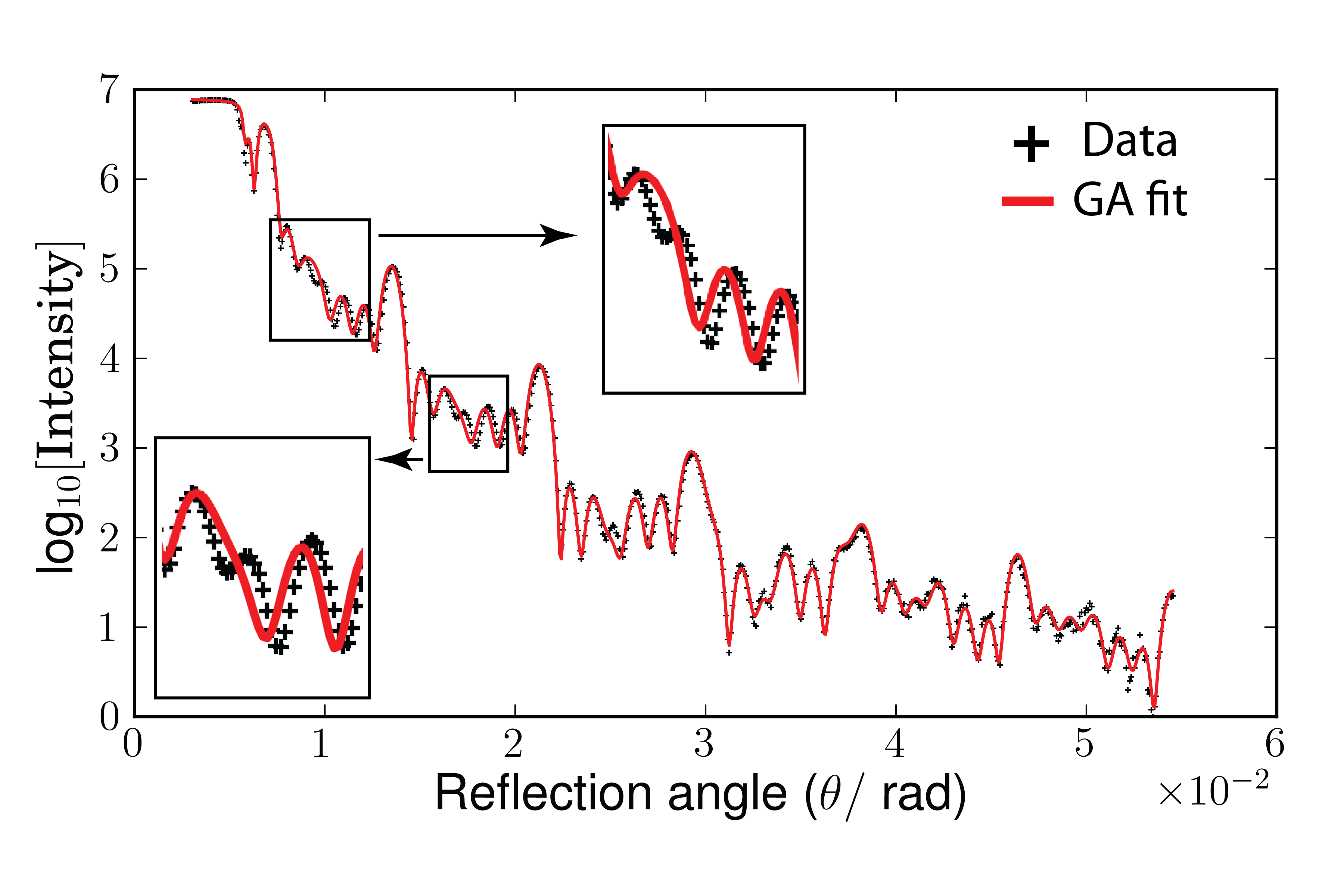}}\\
  b)&\resizebox{\sizeimage\textwidth}{!}
  	{\includegraphics{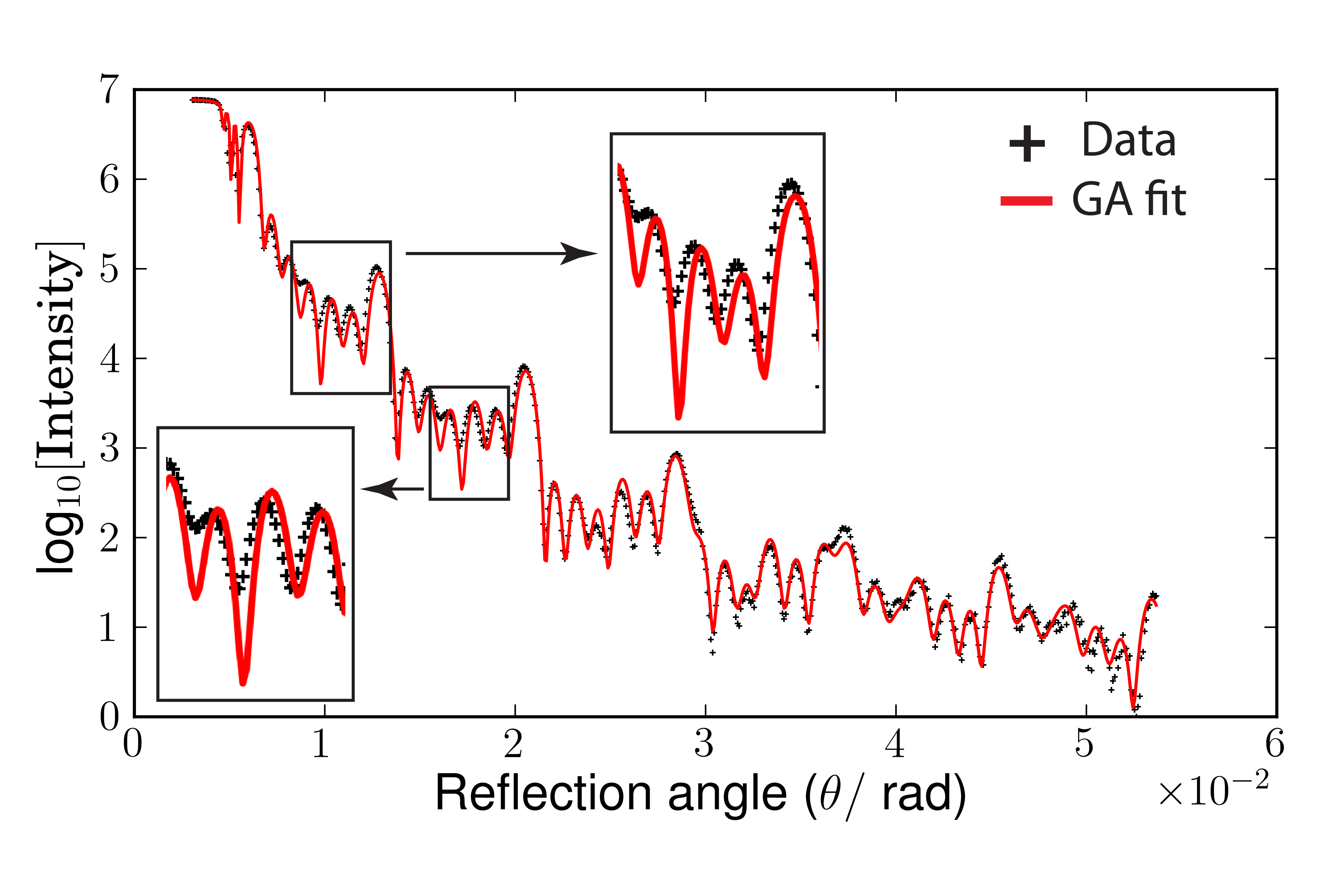}}
  \end{tabular}
  \caption{XRR measurement data and GA model refinement result for data intentionally shifted: a) +0.025$^\circ$ and b) -0.025$^\circ$ $\delta$ in $\omega$.  Data is represented by points with a line representing GA refinement. Inset boxes are magnified views of two regions, showing deviations between refinement and data.}
\label{fig:XRR_+05}
\end{figure}

Figure \ref{fig:XRR_good_fit} shows: a) the original data (aligned) GA refinement, and b) the $\delta = +0.005 ^\circ$ GA refinement.  In both cases, the fit (line) to the data (points) are nearly perfect, with very little to indicate either misaligned or improperly modeled information. In comparison, figure \ref{fig:XRR_+05} shows the results of: a) $\delta= +0.025 ^\circ$ GA refinement and b) $\delta = - 0.025^\circ$ GA refinement.  In both cases, we see substantial deviations between GA (line) and data (points).  Several peaks are missed entirely in both refinements, and oscillation intensities vary substantially between model and data.  This will have a significant impact on the density parameter estimates for these highly misaligned cases.

\begin{figure}
  \resizebox{\sizeimage\textwidth}{!}
  	{\includegraphics{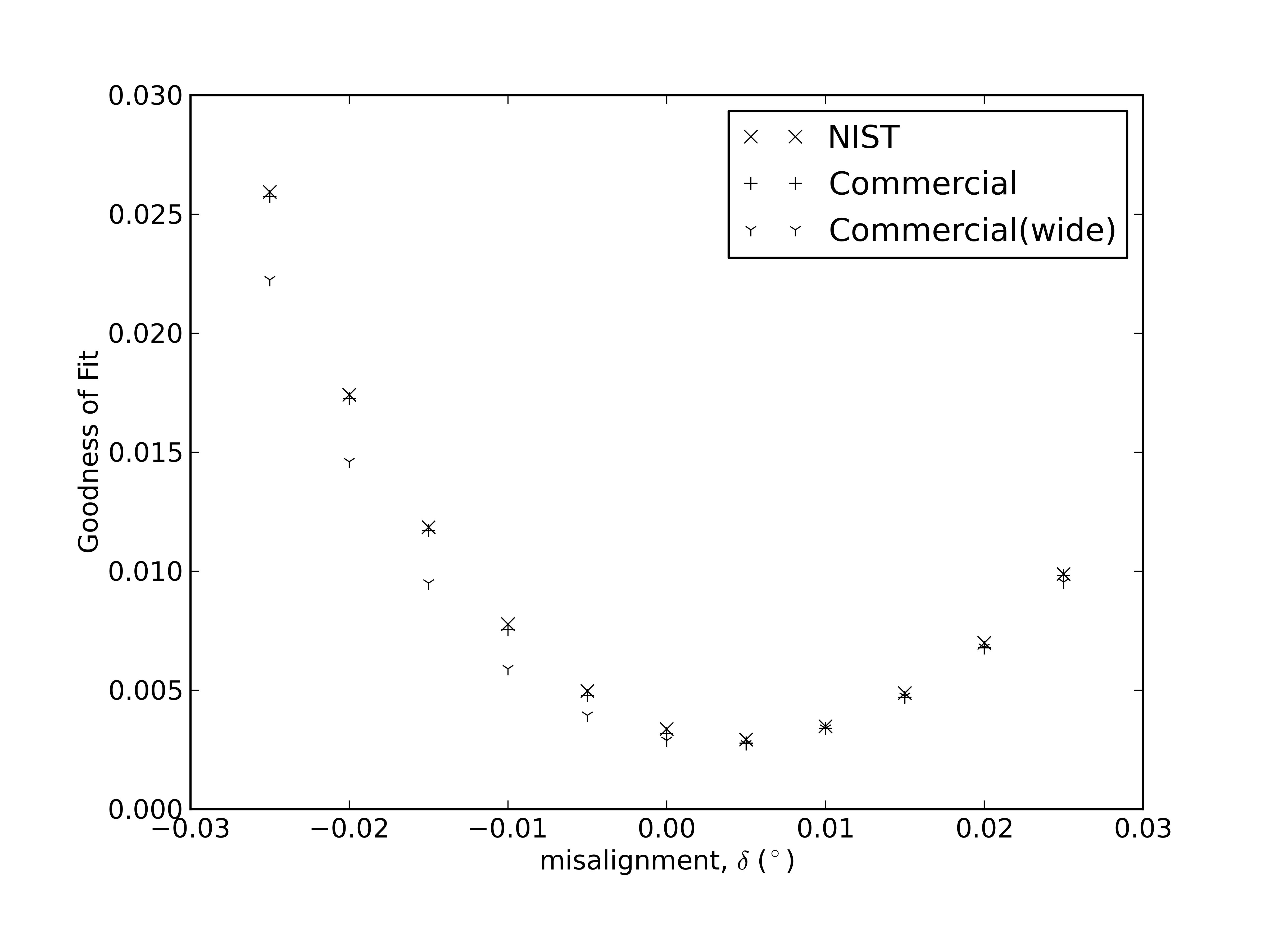}}
  \caption{Goodness-of-fit for the GA refinement of the XRR multilayer model as a function of sample misalignment, $\delta$. NIST developed XRR code is represented by x.  A commercial GA XRR code is shown for comparison with + symbols.}
\label{fig:GOF}
\end{figure}

Figure \ref{fig:GOF} shows the GOF results for the $MSE_{log}$ GA refinements for NIST, commercial, and commercial with wide allowed roughness ranges.   The better agreement between model and data from figure \ref{fig:XRR_good_fit} is clearly seen by a minimum in the range of $\delta = 0$ to $+0.005 ^\circ$.  However, we do see a systematic shift between the GOF results from using Table \ref{table:range} and Table \ref{table:range_wide} (wide).  Wider roughness ranges accommodated more opportunities for surface roughness to exchange with surface layer thickness and provide lower minimum GOFs for the negative offset angle,  $\delta<0$, cases.  This illustrates the difficulty in using GOF as the mechanism for determining optimal alignment.  Bias from allowed parameter ranges can directly affect GOF.

\begin{figure}
  \begin{tabular} {cc}
    a)&\resizebox{\sizeimage\textwidth}{!}
  	{\includegraphics{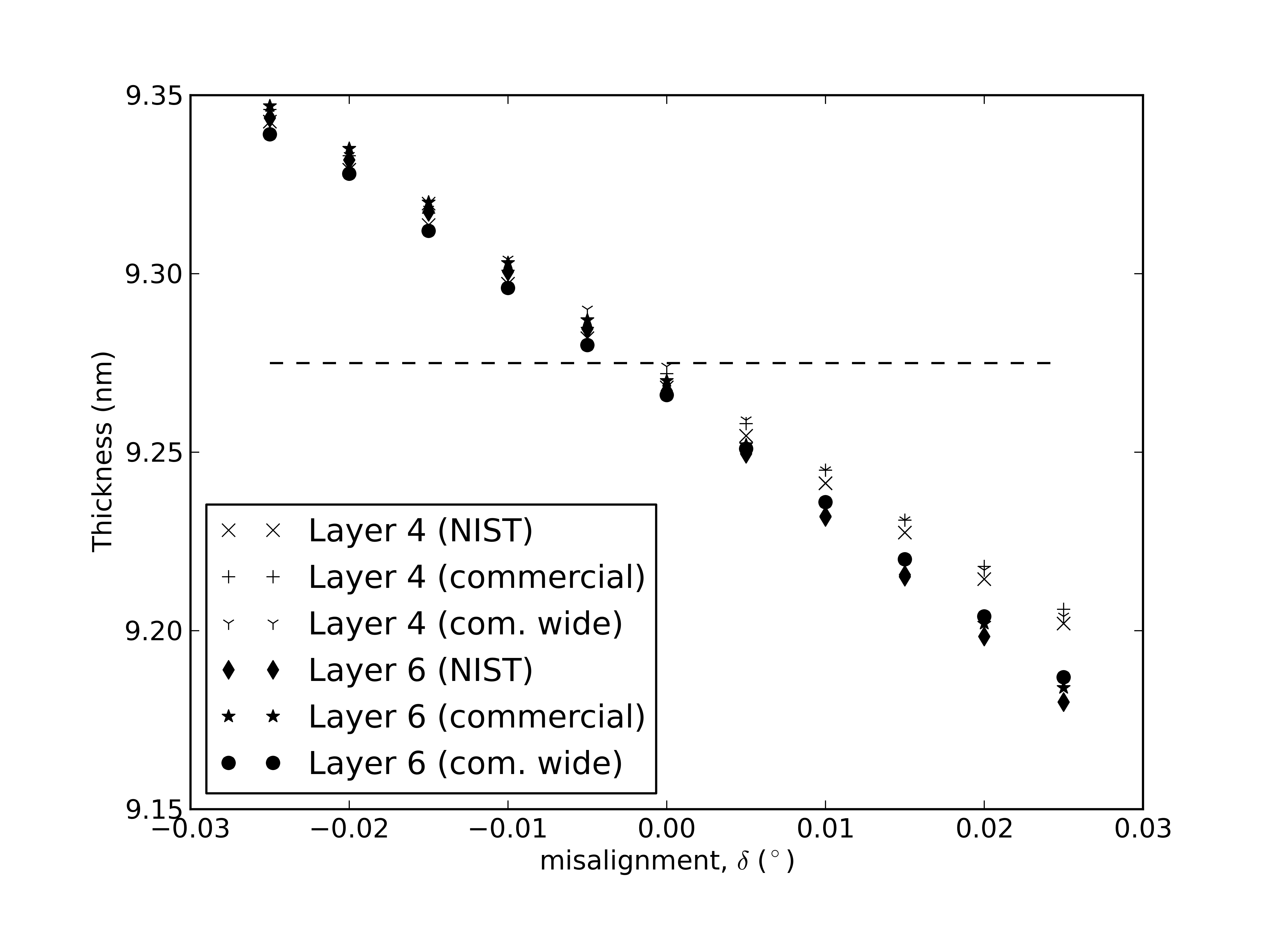}}\\
    b)&\resizebox{\sizeimage\textwidth}{!}
  	{\includegraphics{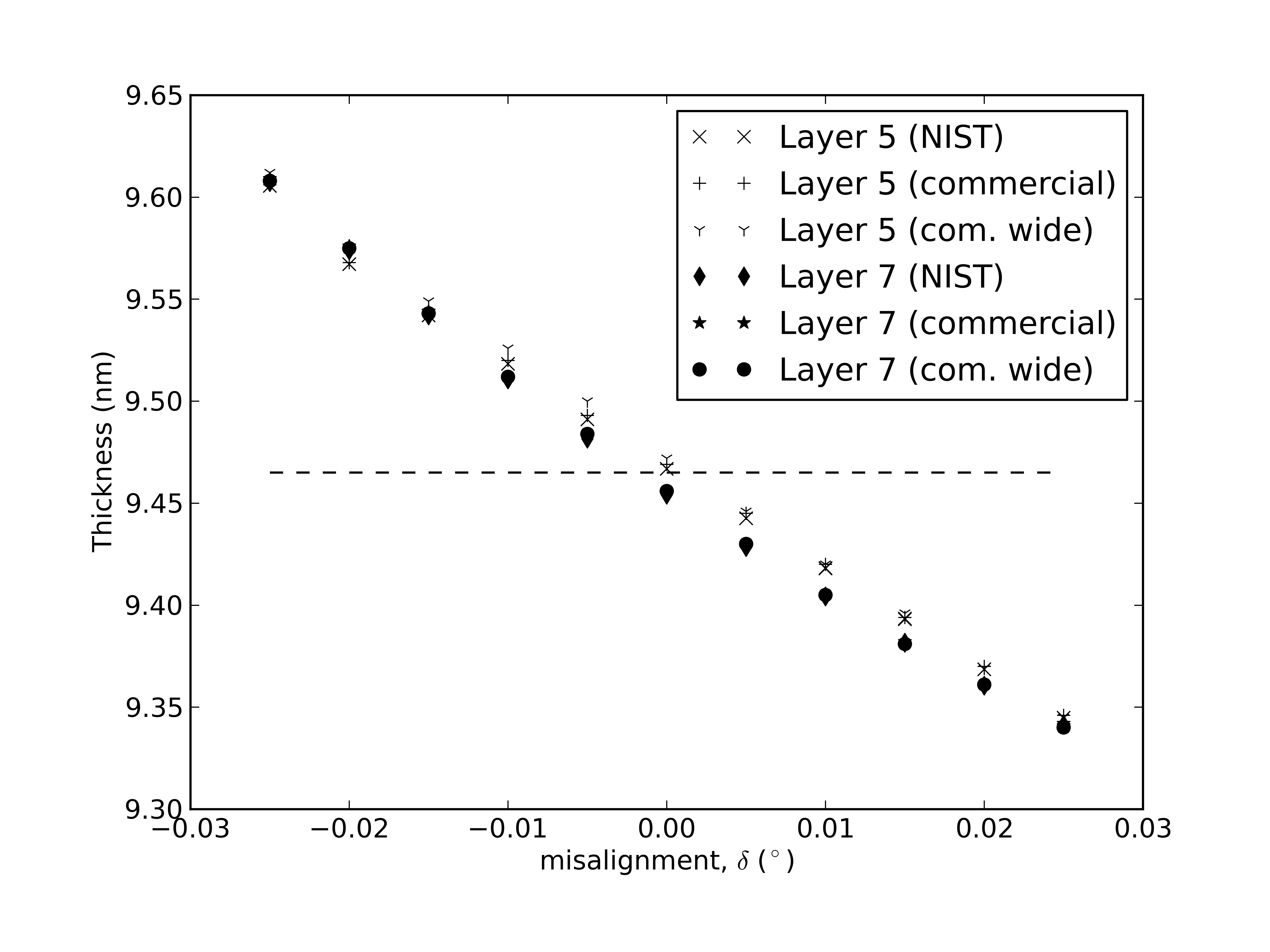}}
  \end{tabular}
  \caption{Thickness determination of: a) GaAs layers, and b) AlAs layers, in multilayer stack, as a function of sample misalignment, $\delta$.  Note a decrease in slope for thickness as a function of sample misalignment.  Dashed line represents the CRM thickness.}
\label{fig:t}
\end{figure}

Evaluation of top layer(s) thickness as a function of $\delta$ led to uninterpretable results containing sharp discontinuities.  Instead, we concentrated on layers deeper into the structure.  Figure \ref{fig:t} shows thickness as a function of $\delta$ for the bottom four buried GaAs and AlAs layers (see Table \ref{table:range} for stack numbering).  All layers show a surprisingly linear relationship suggesting the utility of using a simple linear regression analysis which calculates the quality of the linearity or the coefficient of determination, $R^2$, with regards to calculated thickness and misalignment.  In Table \ref{table:thick_linear} we show these linear regression results for all four layers.  The table shows that sample 1-08 evaluated thickness and intercept show excellent agreement, indicating that the sample was extremely well-aligned.  Note that the $\delta = -0.025 ^\circ$ was omitted in linear refinements due to non-linear behavior in some parameters.  The slope term represents the change in thickness with change in $\delta$.  The negative slope indicates a decrease in thickness when data is misaligned in the $+ \theta^i$ direction.  The $R^2$ values very close to 1 give us very high confidence in linearity.

\begin{figure}
  \begin{tabular} {cc}
    a)&\resizebox{\sizeimage\textwidth}{!}
  	{\includegraphics{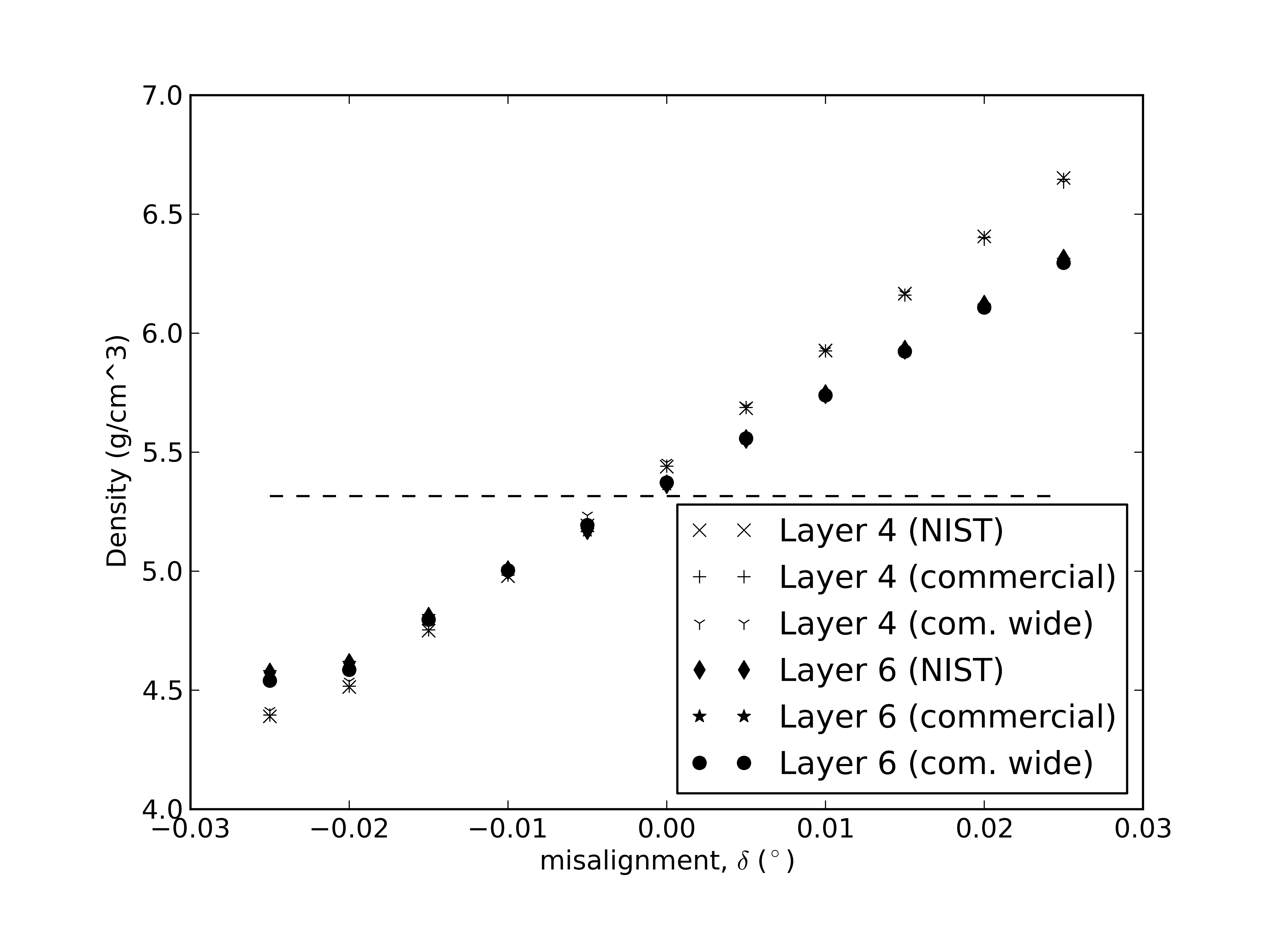}}\\
    b)&\resizebox{\sizeimage\textwidth}{!}
  	{\includegraphics{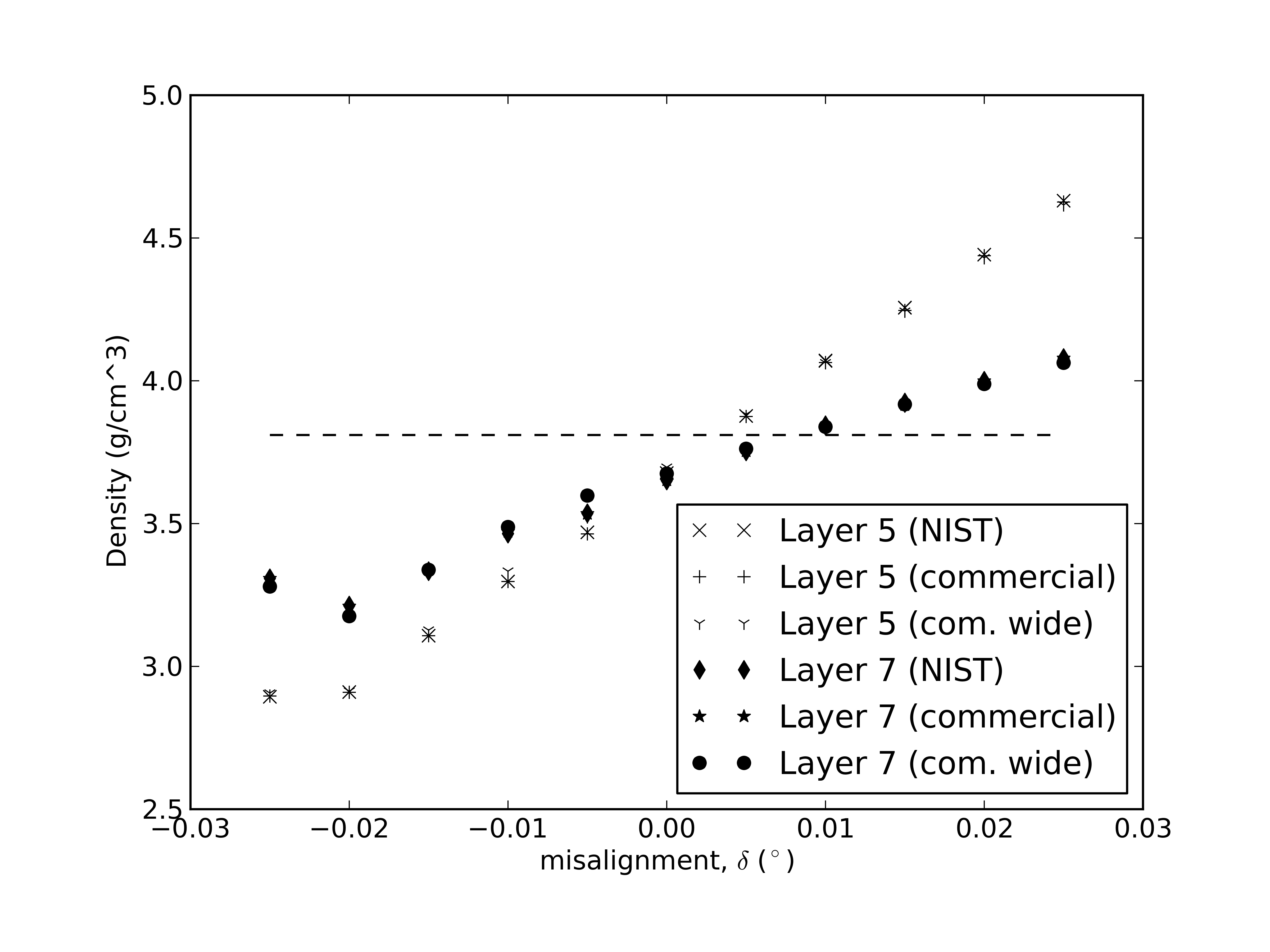}}
  \end{tabular}  \caption{Density determination via genetic algorithm of: a) GaAs layers, and b) AlAs layers, in multilayer stack, as a function of sample misalignment, $\delta$.  Note the variation in the slope of density variance between layers in the stack.  Layers 4 and 5 show larger slopes (more sensitivity to misalignment).}
\label{fig:d}

\end{figure}

In figure \ref{fig:d}, we see density as a function of $\delta$ for the same four buried layers.  Although all of the four layers show highly linear behavior, we see that the slopes vary depending on the layer under study.  There is a trend for higher sensitivity in density determination with layers closer to the surface.  In Table \ref{table:density_linear}, we show both intercept and slope and coefficient of determination for the density variation data.   In this case, we see several key features: First, the intercept densities from the model do not correspond to densities for either bulk AlAs or GaAs.  For GaAs the intercept density is higher, and for AlAs the intercept is lower, than theoretical bulk values.  This could either indicate a bias imposed by other constraints within our structural model, incorrect electron density information within our analysis (i.e., errors in $\delta$ or $\beta$), or it could tell us something about thin film versus bulk density for GaAs and AlAs grown by the molecular beam epitaxy.  There may even be a strain component to this film versus bulk density difference.  All of these possible causes in difference between film and bulk density represent an excellent opportunity for further study. Second, the change in density with respect to misalignment angle is positive and much steeper than the corresponding change in thickness.  We see that density increases with increasing shifts in $\theta^i$ and that density determination through XRR is highly dependent on, and sensitive to, sample alignment.  A small misalignment angle has a pronounced shifting effect on calculated density. This may explain some of the sample-to-sample and measurement-to-measurement variation in density often seen from XRR analysis. The $R^2$ values for density slopes, which demonstrate the linearity of angle versus derived density are very close to 1 (indicating nearly a straight line), except in the case of layer 7 (AlAs layer at the surface of the GaAs substrate) which shows some variation from linearity.



The linear nature of the thickness and density with respect to $\delta$ suggests that first order extrapolations could provide us with a test for systematic errors of sample misalignment in commercial XRR instruments.  Most XRR tools have an automated alignment procedure to validate the specular nature of a measurement prior to collection.  Performing XRR scans with this GaAs/AlAs pre-standard (or an available multilayer standard, such as the NMIJ CRM) for several sample re-mountings could establish the alignment repeatability for the instrument.  In this scenario, some number of data sets (say 10) could be collected and analyzed with commercial GA software, and then the thickness and/or density parameter estimates could be compared to the reference values, to establish any systematic differences in thickness and/or in density, following our determined slope relationships.  As an example, Tables \ref{table:thick_linear} and \ref{table:density_linear} provide a simple alignment recipe for an instrument, if one has the same pre-standard sample available to measure.  A density for Layer 6 of $5.74  \textrm{g cm}^{-3}$ would indicate a misalignment $\delta$ of $ 0.01 ^{\circ}$. If this higher density value was determined consistently over a set of 10 runs, the instrument would be clearly misaligned and could be offset accordingly.  This approach would work on any tool, provided the same energy (Cu K$_\alpha$), same angular range, and same model parameter ranges (see Table \ref{table:range_wide}) were used in the refinement, and that a refinement software could achieve a true global minimum for each analysis.  Radically different beam conditioning optics may impact this type of analysis, and sample height errors can also have consequences.  A more rigorous study is needed to establish the second order impacts of these additional instrumental effects.

\section{Conclusions}

This paper illustrates the merits of using a Certified Reference Material, in this case a multilayer AlAs/GaAs CRM from NMIJ, in the assessment of sample alignment for the X-ray Reflectometry method.  Sample alignment information can be better inferred through thickness (and possibly density) variation, than through more traditional GOF minimization.  Over a reasonable range of sample misalignments ($\delta \pm 0.02^\circ$), both thickness and density vary linearl for the buried layers, allowing for direct extrapolation of misalignment effects.  A combined strategy of assessing both thickness and density from buried layers of the multilayer structure may yield the best alignment assessment ($U\{\delta\} < 0.005^\circ$).  This combined approach needs further study, as density parameters have not been certified.  Note that this technique, by its nature, relies on the availability and validity of thickness and/or density certified parameters from a traceable reference material.  By comparing systematic bias in these parameters, it is possible to assess potential misalignment errors caused by mounting and stage/sample alignment approaches on a commercial reflectometer.  This is very important in the field, as sample alignment is often the dominant source of uncertainty in XRR measurement for high quality films.

To date, these studies have only been developed using a single data set.  Future work in this area will involve the automation of this alignment testing procedure to allow for repeatability testing of these results from multiple measurements and multiple instruments using the same specimen (CRM) to explore the limits of this approach.


\section*{References}

\bibliographystyle{unsrt}   

\bibliography{07_MRS_windover_mod,2009_frontiers_xrr,library}

\newpage


\begin{table}
\caption{GA parameter ranges for multilayer refinement.}
\label{table:range}
\begin{tabular}{ccccc}
 \br
Layer 
&Material
& $t / \textrm{nm}$
 & $\sigma / \textrm{nm}$
 & $\rho / \textrm{g cm}^{-3}$
 \\
 \mr
1& Al$_2$O$_3$ & 1.0-5.0 & 0.1-0.3 & 1.0-5.0 \\
2&GaAs& 6.0-11.0 & 0.3-0.5 & 2.66-7.98\\
3&AlAs & 9.0-10.0 & 0.3-0.5 & 1.91-5.72 \\
4&GaAs& 9.0-10.0 & 0.3-0.5 & 2.66-7.98\\
5&AlAs & 9.0-10.0 & 0.3-0.5 & 1.91-5.72 \\
6&GaAs& 9.0-10.0 & 0.3-0.5 & 2.66-7.98\\
7&AlAs & 9.0-10.0 & 0.3-0.5 & 1.91-5.72 \\
8&GaAs& -- & 0.3-0.5 & 5.316\\
 \br
\end{tabular}
\end{table}

\begin{table}
\caption{GA with wide roughness ranges for multilayer refinement.}
\label{table:range_wide}
\begin{tabular}{ccccc}
 \br
Layer 
&Material
& $t / \textrm{nm}$
 & $\sigma / \textrm{nm}$
 & $\rho / \textrm{g cm}^{-3}$
 \\
 \mr
1& Al$_2$O$_3$ & 1.0-5.0 & 0.1-2.0 & 1.0-5.0 \\
2&GaAs& 6.0-11.0 & 0.1-1.0 & 2.66-7.98\\
3&AlAs & 9.0-10.0 & 0.1-1.0 & 1.91-5.72 \\
4&GaAs& 9.0-10.0 & 0.1-1.0 & 2.66-7.98\\
5&AlAs & 9.0-10.0 & 0.1-1.0 & 1.91-5.72 \\
6&GaAs& 9.0-10.0 & 0.1-1.0 & 2.66-7.98\\
7&AlAs & 9.0-10.0 & 0.1-1.0 & 1.91-5.72 \\
8&GaAs& -- & 0.1-1.0 & 5.316\\
 \br
\end{tabular}

\end{table}

\begin{table}
\caption{Linear fit to thickness in layers from misaligned data.}
\label{table:thick_linear}
\begin{tabular}{cccccc}
 \br
Layer 
&Material
& Sample
& Intercept
& Slope
& coefficient of
\\

&
&1-08
& 
& $m$/
& determination
\\

&
& $t / \textrm{nm}$
& $t / \textrm{nm}$
& $\textrm{nm}$ $(^\circ)^{-1}$
& $\textrm{R}^{2}$  
\\
 \mr
1& Al$_2$O$_3$ & NA \\
2&GaAs& NA \\
3&AlAs & NA \\
4&GaAs& 9.281 & 9.272 & -2.845 & 0.996 \\
5&AlAs & 9.468 & 9.469 & -4.990 & 0.999 \\
6&GaAs& 9.269 & 9.267 & -3.311 & 0.997 \\
7&AlAs & 9.458 & 9.461 & -5.193 & 0.994 \\
8&GaAs& NA\\
 \br
\end{tabular}
\end{table}

\begin{table}
\caption{Linear fit to density in layers from misaligned data.}
\label{table:density_linear}
\begin{tabular}{cccccc}
 \br
Layer 
&Material
& ``Bulk''
& Intercept
& Slope
& coefficient of
\\

&
&
& 
&$m$/
& determination
\\
&
& $\rho / \textrm{nm}$
& $\rho / \textrm{nm}$
&  $\textrm{nm}$ $(^\circ)^{-1}$
& $\textrm{R}^{2}$  
\\
 \mr
1& Al$_2$O$_3$ & NA \\
2&GaAs&  NA \\
3&AlAs & NA \\
4&GaAs& 5.316 & 5.455 & 47.24 & 0.9997 \\
5&AlAs & 3.81 & 3.680 & 38.05 & 0.9995 \\
6&GaAs& 5.316 & 5.367 & 37.53 & 0.9998 \\
7&AlAs & 3.81 & 3.633 & 19.01 & 0.989 \\
8&GaAs& NA\\
 \br
\end{tabular}
\end{table}

\end{document}